\documentclass[preprint,12pt]{elsarticle}

\usepackage{amssymb}
\usepackage{amsthm}

\usepackage{amsmath}

\journal{Physica D}
\begin{document}

\begin{frontmatter}



\title{Periodic Orbit Scar in Propagation of Wavepacket}


\author{Mitsuyoshi Tomiya}

%\address{Faculty of Science and Technology, Seikei University, 
%Kichijyoji-Kitamachi 3-3-1, Musashino-shi, Tokyo, 180-8633, JAPAN}

\author{ Shoichi Sakamoto}

\address{Faculty of Science and Technology, Seikei University, 
Kichijyoji-Kitamachi 3-3-1, Musashino-shi, Tokyo, 180-8633, JAPAN}

\author{Eric J. Heller}

\address{Department of Physics, Harvard University, 
Cambridge, Massachussetts 02138, USA}

\begin{abstract}
This study analyzed the scar-like localization in the time-average of a time-evolving wavepacket on the desymmetrized stadium billiard.  When a wave-packet is launched along the orbits, it emerges on classical unstable periodic orbits as a scar in the stationary states.
This localization along the periodic orbit is clarified through the semiclassical approximation.  It  essentially originates from the same mechanism of a scar in stationary states: the piling up of the contribution from the classical actions of multiply repeated passes on a primitive periodic orbit.
To create this enhancement, several states are required in the energy range, which is determined by the initial wavepacket.
%  Most of these states have dominant contribution on the specific periodic orbit.
\end{abstract}


\end{frontmatter}


%\section{}
%\label{}





\section{\label{sec:level1}Introduction}

This study investigates the localization in the time-average of the absolute squares of the time-evolving wave function on the desymmetrized stadium billiard that occurs after the Gaussian wavepacket is launched as the initial state. 
In chaotic billiards like a stadium, the nodal patterns of stationary states with unique characteristics were discovered approximately three decades ago \cite{heller}.   The patterns often have a unique enhancement along classical unstable periodic orbits.  Such a phenomenon is called scar in quantum stationary states of a finite chaotic region.  The eigen states with scars are called scar states.  In contrast, in integrable billiards, the nodal patterns are essentially repetitive and synthetic. 
The eigen states are a genuine quantum mechanical concept, whereas the periodic orbits are apparently classical mechanical objects.  The scar state is an important discovery expressing a providential quantum-classical correspondence.

\begin{figure}
\includegraphics[width=13cm]{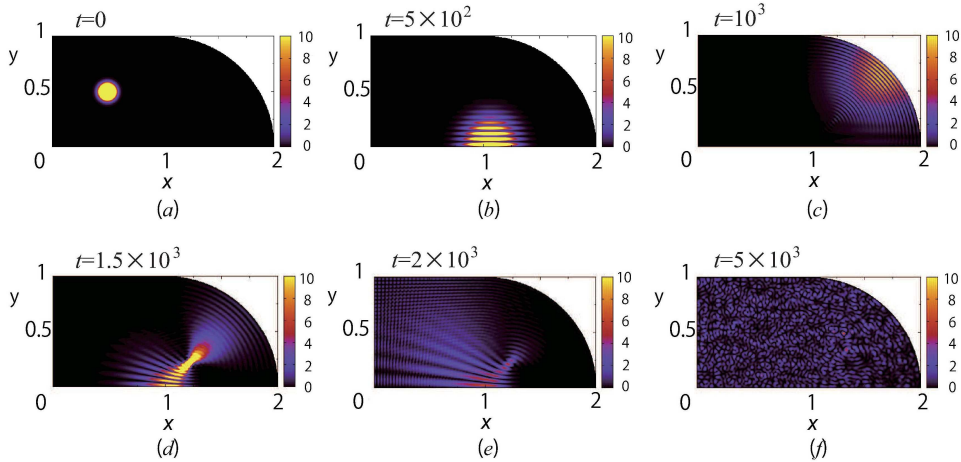}
\caption{ 
(a)-(f) Illustrations of time-evolution of the Gaussian wavepacket in a desymmetrized stadium billiard.  The $x$-coordinate is set along the bottom line of the stadium, and the $y$-coordinate is on the left straight boundary.  Thus, the origin of the coordinate is located on the left bottom corner.   
The wavepacket is launched from $\mathbf{r}_0=(1/2, 1/2)$  and begins travelling with the launching angle $\theta = -\pi/4$(a), which is defined in the counterclockwise direction from the direction of the $x$-axis,  
$|\mathbf{p}_0| = 250$, and $\sigma_0=0.15$.  
The orbit corresponds to periodic orbit No.7 in \cite{Bogomolny}.
After approximately $t=5 \times 10^3$, the wave function almost defuses all over the stadium(f). 
\label{fig.1}}
\end{figure}

A semiclassical approximation emerged as a powerful tool to clarify scar states in quantum systems along the classical unstable periodic orbits. This method has been used to construct theories of scars in coordinate space \cite{ Bogomolny} and phase space \cite{LesHouches, Berry-Wignerdist}; they successfully clarify the contribution of the periodic orbits to the scar states. 
Both theories discuss the scars in energy dependence  because the scars first are discovered in the eigen states.  
Bogomolny \cite{ Bogomolny} proposed a Green's function in terms of actions of classical periodic orbits to expose the periodic orbits as the origins of the scar in the coordinate space.  Berry's theory \cite{ Berry-Wignerdist} utilizes the Wigner function under approximation in the phase space to clarify the cause of the scars. 
In particular, Heller's lecture \cite{LesHouches} revealed the dynamical properties of scars, stating that the time-evolving wavepackets propagate near the periodic orbits. 
Especially, the Heller group focused on homoclinic orbits and the return of the Gaussian wavepacket
 to the neighborhood of its launching point in finite regions. In addition, they realized the importance of the autocorrelation function and its Fourier counter part: the weighted spectrum \cite{heller2, TH, gaussian, KH-LinearNonlinear, KH-shorttime}.

Finally, the enhancement or localization in the time-average of the time-evolving wavepacket was discovered \cite{ourpaper, ourpaper2}.  
In this study, it is called as the ``dynamical scar".  It has a distinctly close relation to scar states because it also emerges along a periodic orbit \cite{prep}.  In this study, the scar states are shown to heavily contribute to the dynamical states.  The window function \cite{St} for the semiclassical approximation to describe the enhancement is derived from the weighted power spectrum.

However, it is known that reflection symmetries of billiard's shape sometimes prevents the detection of its genuine chaotic characteristics.  To remove the discrete symmetries, we studied the localization in a desymmetrized $2\times4$ stadium billiard \cite{Bunimovich}.   The desymmetrization eliminates the two discrete mirror symmetries of the full stadium shape and makes the chaotic properties more evident. We use Table I in Ref.\cite{Bogomolny} to distinguish the periodic orbits; however, the table is for a full stadium, and not for a desymmetrized stadium.  Therefore, it should be used with caution.
If the periodic orbits pass over the horizontal and vertical axes of the symmetries, they may have to be folded at the crossing points for the desymmetrized stadium (cf. Fig.2,3).

\section{Gaussian wavepacket as a probe for dynamical properties}

The time-dependent Schr\"{o}dinger equation
\begin{equation}
i \hbar \frac{\partial \Psi}{\partial t}=  - \frac{\hbar^2}{2m} \nabla ^2  \Psi + V  \Psi  
\end{equation}
governs dynamical properties of quantum systems.  By adopting the quarter of the $2 \times 4$ stadium (FIG.1$-$3)  as the 2D chaotic finite structure, the potential is simply set to $V=0$ inside the billiard and $V=\infty$ outside.      

The Gaussian wavepacket is a conventional tool used for elucidating the time-evolution of quantum states \cite{TH, gaussian, KH-LinearNonlinear, KH-shorttime}. 
It has been one of the fundamental quantum objects since the early stage of quantum mechanics. 
Its initial form in a 2D region is
\begin{equation}
\Psi_0 (\mathbf{r}) = \frac{1}{ \sigma_0 \sqrt{\pi} }  
exp \left[ \frac{i}{\hbar} \mathbf{p}_0 ( \mathbf{r} - \mathbf{r}_0 ) - \frac{(\mathbf{r} - \mathbf{r}_0)^2}{2 {\sigma_0}^2} \right] ,
\end{equation}
where  $\mathbf{r}=(x, y)$ is a point inside the nanostructure,   
$\mathbf{r_0}=(x_0, y_0)$ is the initial location of the center of the wavepacket, 
and  ${\mathbf{p}_0}=(p_{0x},p_{0y})$ is the packet's initial momentum. 
The standard deviation of the Gaussian packet $\sigma_0$ determines its size.

\begin{figure}
\includegraphics[width=13cm]{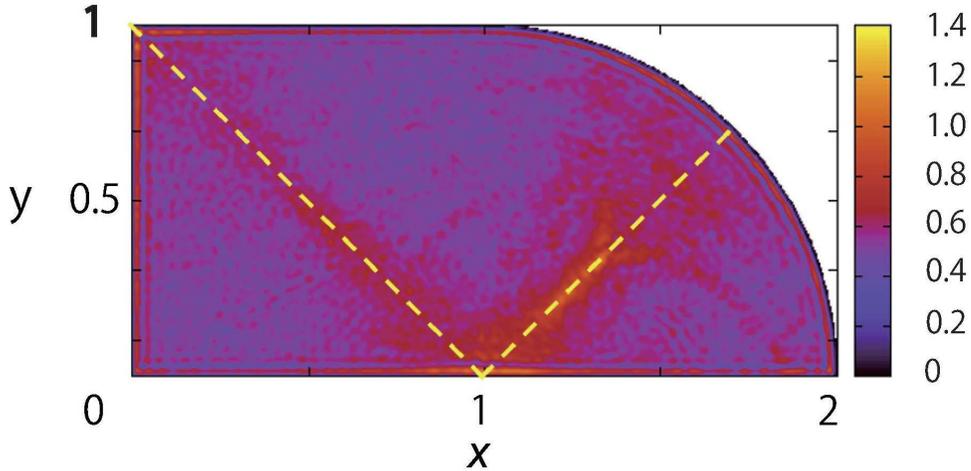}
\caption{ 
The time-average of the evolving wavepacket $A(\mathbf{r})$ in FIG.1.
The weak concentration appears along the broken yellow lines which represent the corresponding unstable periodic orbit.  
It shows the shape of the desymmetrized orbit No.7 in \cite{Bogomolny}.
\label{fig.2}}
\end{figure}

If the Gaussian wavepacket is placed in a flat infinite space, 
it travels as a bunch with the initial velocity of the center of the wavepacket $\mathbf{v}_0=\mathbf{p}_0 / m $.
The absolute value of the wavepacket shows that its shape is always Gaussian;  
however, its size increases as  $|\sigma(t)|=\sigma_0 \sqrt{1+ \left( \frac{\hbar t}{{m}{\sigma_0}^2} \right)^2 } $.  If time is sufficiently long, $\sigma(t) \approx \frac{\hbar}{m \sigma_{0} } t $.   

In this study, the wavepacket travels in the finite region, 
and repeated reflections on the boundary eventually diffuse it all around the billiard (Fig.1; cf. \cite{ourpaper, ourpaper2}).
Initially it behaves like a bunch of viscous liquid. 
The travelling wavepacket then gradually and progressively shows less specific texture.
Finally, in chaotic billiards, the snapshots of wave function ripple all over the billiard with irregular granular pattern.    
On the contrary, the autocorrelation function has surprisingly already revealed long time recurrence \cite{gaussian}.  Moreover, this obliquely implies the localization on the periodic orbit.

\section{Dynamical scar}

One of the most fundamental concepts in quantum physics is the use of the absolute square of the wave function to derive any physical properties; usually its time average is important to investigate a quantum effect. 
Therefore, the time-average of the absolute square of the wave function is as follows:   
\begin{equation}
A_{T}(\mathbf{r})= \frac{1}{T} \int_0^T |\Psi(\mathbf{r},t)|^2 dt . 
\end{equation}
This is an appropriate tool to detect the localization that is concerned.  Here, $T$ expresses the time required to measure the time-average.

For numerical calculation, it is discretized
as
\begin{equation}
A_{T}(\mathbf{r}_{i})=\frac{1}{N_t}\sum_{j=0}^{N_t} | \Psi( \mathbf{r}_{i}, t_{j} )|^{2} ,
\end{equation} 
on the mesh points $\mathbf{r}_i=(x_i,y_i)$,
and the integration over time is the summation over
the discretized times $t_j = j \Delta t  $, where $\Delta t$ is a time step. The summation must then be divided by the integer $N_t$ representing the number of whole time steps and apparently $T=N_t \Delta t$. 
In this study, the natural units $\hbar=m=1$ are always applied for actual numerical evaluation. The time step is set at $\Delta t= 2.5 \times 10^{-2}  $,
$T=9 \times 10^4$, or $N_t = 3.6 \times 10^6$,
and the lattice constant is 0.2.
A typical example of calculated $A_T$ is presented in Fig.2. 
The time-average expresses clear localization along unstable periodic orbits despite no specific patterns in the snapshots of the wavepackets (e.g. Fig.1(f)).

It is apparently similar to the scars of a stationary wave function \cite{heller}. Furthermore, different launching conditions exibit the same phenomena on various periodic orbits, as shown in Fig.3 (also see \cite{prep}).  The enhancement appears clearly around the periodic orbit if the initial location of the center of the wavepacket and its velocity are on and along the orbit.  
These are referred to as ``dynamical scars" to distinguish them from the scar states in stationary eigen states.
These are an enhancement in the time-average of time-dependent wave function.   

Any states in quantum systems can be expanded using these eigenfunctions as
\begin{equation}
\Psi(\mathbf{r},t)=\sum_n c_n \psi_n (\mathbf{r},t)=\sum_n c_n \phi_n (\mathbf{r}) \exp(- \frac{i}{\hbar} E_n t),
\end{equation}   
where $\psi_n (\mathbf{r},t) = \phi_n (\mathbf{r}) \exp(- \frac{i}{\hbar} E_n t)$ is the $n$-th eigen state of the system with energy $E_n$.
The expansion coeffient $c_n$ must satisfy the condition $\sum_n |c_n|^2=1$.
In this study, the initial state is set  $\Psi(\mathbf{r},t=0)=\Psi_0(\mathbf{r})$.
The expansion coefficient $c_n$ can be determined using the initial wavepacket $\Psi_0$ as 
\begin{equation}
c_n = \int \phi_n^* \Psi_0 (\mathbf{r}) d \mathbf{r} .
\end{equation}
Moreover, the expansion can be used to elucidate the time-average of $|\Psi(\mathbf{r},t)|^2$ as
\begin{align}
A(\mathbf{r})=& \lim_{T \to \infty}A_{T}(\mathbf{r}) 
= \lim_{T \to \infty} \frac{1}{T}  \int_0^T |\Psi(\mathbf{r},t)|^2 dt \nonumber \\
=& \lim_{T \to \infty} \frac{1}{T} \int_0 ^T \biggl[ \sum_n |c_n|^2 |\phi_n (\mathbf{r})|^2 + \sum_{n \neq m} c_m^* c_n \phi_m^* \phi_n exp{ \Bigl\{ \frac{i}{\hbar} (E_m - E_n) t 
\Bigr\} }  \biggr] dt \nonumber  \\
=& \sum_n |c_n|^2 |\phi_n (\mathbf{r})|^2   , 
\end{align}
assuming $E_n \neq E_m$, if $n \neq m$.  
In other words, by Eq.(7), if the coefficients $c_n$ of the scar eigen states on the same periodic orbit have dominantly larger values, ``dynamical scars" of the periodic orbits are observed in the time-average $A(\mathbf{r}$) \cite{ourpaper,ourpaper2,prep}.  

Therefore, at least theoretically, the time-average (7) can be written in energy integration as follows:
\begin{equation}
A(\mathbf{r})
= \int \sum_n  |c_n|^2  |\phi_n (\mathbf{r})|^2 \delta (E-E_n) dE  .
\end{equation}
However, the Dirac delta function must be treated carefully to allow comparison of numerical results and experimental data.  The behavior of the delta functions is often smoothed by the limitation of the precision of numerical calculation and experimental measurement.

 Eq.(8) can be considered as the summation of the related wave fuctions and the specific contribution weight that closely correspond to the weighted spectrum because it includes the factor $|c_n|^2$.  In numerical calculation, the weighted spectrum would be smoothed by the numerical discretization and the precision of the calculation.  The Dirac delta function could be replaced with a smoothed function.

\begin{figure}
\includegraphics[width=13cm]{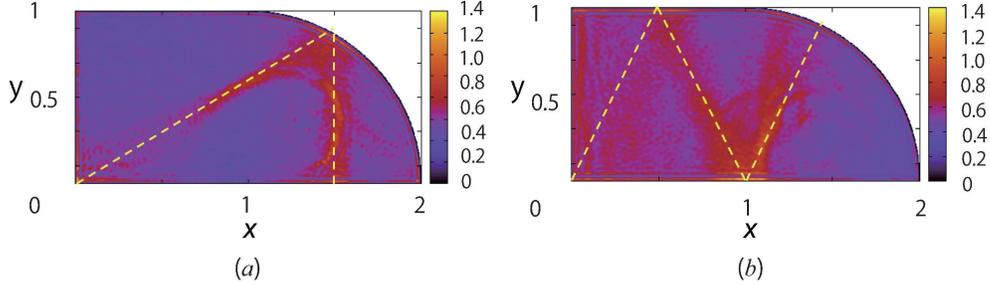}
\caption{The time-averages of the evolving wavepackets $A(\mathbf{r})$ in stadium billiard with different initial conditions. In both cases for the initial Gaussian wavepackets, $|\mathbf{p}_0| = 250$ and $\sigma_0=0.15$.  (a) The wavepacket is launched from $(1/2, \sqrt{3}/6)$ and its lanuching angle is $\theta=\pi/6$. This shows the shape of the desymmetrized orbit No.12 in \cite{Bogomolny}. (b) The wavepacket launched from  $(1/4, 1/2)$ has an angle defined as $tan \theta=2$.  This corresponds to orbit No.14 in \cite{Bogomolny}. The launching angles are defined as those in Fig.1. The broken yellow lines correspond to the classical unstable periodic orbits.
\label{fig.3}}
\end{figure} 

\section{Window function}

The correlation function between the travelling wavepacket (5) and initial state (2) $C_0(t) = \int \Psi_0^* (\mathbf{r}) \Psi(\mathbf{r},t) d\mathbf{r}^2$ closely relates to the weighted spectrum.  The autocorrelation function is expressed by the eigenfunction expansion (5) as
\begin{align}
C_0(t) &= \int \Psi_0^*(\mathbf{r}) \Psi(\mathbf{r},t)  d^2 \mathbf{r}                      \nonumber  \\ 
 &= \int (\sum_m c_m^* \phi_m^*)  ( \sum_n c_n \phi_n e^{-\frac{i}{\hbar}E_n t}) d^2 \mathbf{r}  \nonumber  \\ 
&= \sum_n |c_n|^2 e^{-\frac{i}{\hbar}E_n t}  .
\end{align}
The weighted spectrum can be defined through its Fourier transform as
\begin{align}
\tilde{C}_0 (E) &=\frac{1}{2\pi} \int_{-\infty}^{\infty} C_0(t) e^{\frac{i}{\hbar}Et}dt      \nonumber \\
&=\frac{1}{2\pi} \int_{-\infty}^{\infty} \sum_n |c_n|^2 e^{\frac{i}{\hbar}(E-E_n)t}dt      \nonumber \\
&= \hbar \sum_n |c_n|^2 \delta(E-E_n)                                                              \nonumber \\
&= \hbar P(E).  
\end{align}
This only represents the bare weighted power spectrum $P(E)= \sum_n |c_n|^2 \delta(E-E_n)$ with the Planck constant.
 
The smoothed version of the weighted spectrum and the Green's function introduce a neat form of the time-average.  The smoothed weighted spectrum function (SWSF) can be written as
\begin{equation}
P_\epsilon(E)=\sum_{n} |c_n|^2 \delta_\epsilon (E-E_n) .
\end{equation}
In addition, we have $\lim _ {\epsilon \rightarrow 0} P_{\epsilon}(E) =P(E)$.
 When $\epsilon$ becomes infinitesimal, $\lim_{\epsilon \rightarrow 0} \delta_\epsilon (x) = \delta(x)$.
Here, the Lorentzian form of the smoothed version delta function is introduced as 
\begin{equation}
\delta_\epsilon (E-E_n) = \frac{\epsilon}{\pi}  \frac{1}{(E-E_n)^2 + \epsilon^2} .
\end{equation}
Realistic systems have finite precision and always show errors because of numerical applications, limit of measurement, etc.  
Owing to these inevitable limitations of the systems, the Dirac delta functions are replaced by some finite regular functions.  Its infinity and singular behavior cannot be recreated exactly in a computation; they seem very large but are finite, and are singular-like; however, the peaks are not numerically infinite.  The width of the Lorenzian $\epsilon$ would be the order of the mean level spacing $\overline{\Delta E}$ under such limitation  because much finer energy difference would not be distinguishable.  The replacement is allowed, considering the width of  the  Lorentzian $\epsilon$ should be equal or larger than the order of the mean level spacing $\overline{\Delta E}$.
By using this expression, the smoothed Green's function
\begin{equation}
\rm{Im}G_\epsilon (\mathbf{r},\mathbf{r};E)=-\pi \sum_n |\phi_n (\mathbf{r})|^2  \delta_\epsilon (E-E_n) 
\end{equation}
is also introduced.

Under such circumstances, a square of the delta functions can be treated using Berry's method \cite{Berry85}.  The smoothed delta function (12) has a remarkable property: 
\begin{equation}
\bar{\delta_\epsilon} (E-E_n) =2 \pi \epsilon [\delta_\epsilon (E-E_n)]^2 = \frac{2\epsilon^3}{\pi} \frac{1}{ \{(E-E_n)^2+{\epsilon}^2 \}^2},
\end{equation} 
where $\bar{\delta_\epsilon} (E-E_n)$ is another version of the smoothed delta function $\lim_{\epsilon \rightarrow 0}\bar{\delta_\epsilon} (x) = \delta(x)$.
 Next, we use an alternative practical version of the time-average
\begin{equation}
A_{\epsilon}(\mathbf{r})
= \int \sum_n  |c_n|^2  |\phi_n (\mathbf{r})|^2 \bar{\delta_\epsilon} (E-E_n) dE  .
\end{equation}
The original time-average $A$ is in the limit $A(\mathbf{r}) =  \lim_{\epsilon \rightarrow 0} A_{\epsilon} (\mathbf{r})$.

By multiplying the two terms (11) and (13), we obtain
\begin{align}
P_\epsilon(E)& \rm{Im} \it{G}_\epsilon(\mathbf{r},\mathbf{r};E )       \nonumber  \\
&=\sum_{n} |c_n|^2  \delta_\epsilon(E-E_n)
\bigl \{ -\pi \sum_{n'}|\phi_{n'}(\mathbf{r})|^2 \delta_\epsilon(E-E_{n'}) \bigr\}   \nonumber  \\
&=-\pi \sum_{n,n'} |c_n|^2 |\phi_{n'}(\mathbf{r})|^2   \delta_\epsilon(E-E_n) \delta_\epsilon(E-E_{n'})   \nonumber  \\
&=  -\pi \sum_{n} |c_n|^2 |\phi_{n}(\mathbf{r})|^2 \left[ {\delta_\epsilon}(E-E_n) \right]^2  \nonumber  \\
&=  \frac{-1}{2\epsilon} \sum_{n} |c_n|^2 |\phi_{n}(\mathbf{r})|^2  \bar{\delta_\epsilon}(E-E_n) .   
\end{align}
Here Eq.(14) is also applied for this deformation.
Finally, Eq.(16) is used to provide the following expression for the time-average by using the  Green's function
\begin{align}
A_{\epsilon}(\mathbf{r})
&=-2\epsilon\int_{-\infty}^{\infty} P_{\epsilon}(E) \rm{Im} \it{G}_{\epsilon} (\mathbf{r},\mathbf{r};E) dE  \nonumber \\
&= \int_{-\infty}^{\infty} w(E) \rm{Im} \it{G}_{\epsilon} (\mathbf{r},\mathbf{r};E) dE,
\end{align}
where  the window function $w(E)$ is introduced \cite{St}  through SWSF (11) as 
\begin{equation}
w(E)=-2 \epsilon P_{\epsilon}(E) = - \frac{2\epsilon}{\hbar}  \tilde{C_0}(E).
\end{equation}
In other words, $w(E)$ is the weight for the integration over the energy region to evaluate the time-average $A_{\epsilon} (E)$ from tne imaginary part of the smoothed Green's function (13).  This is the specific quantum phenomenon that is focused upon in this study.  It determines where the window should be transparent in the energy spectrum.

\begin{figure}
\includegraphics[width=13cm]{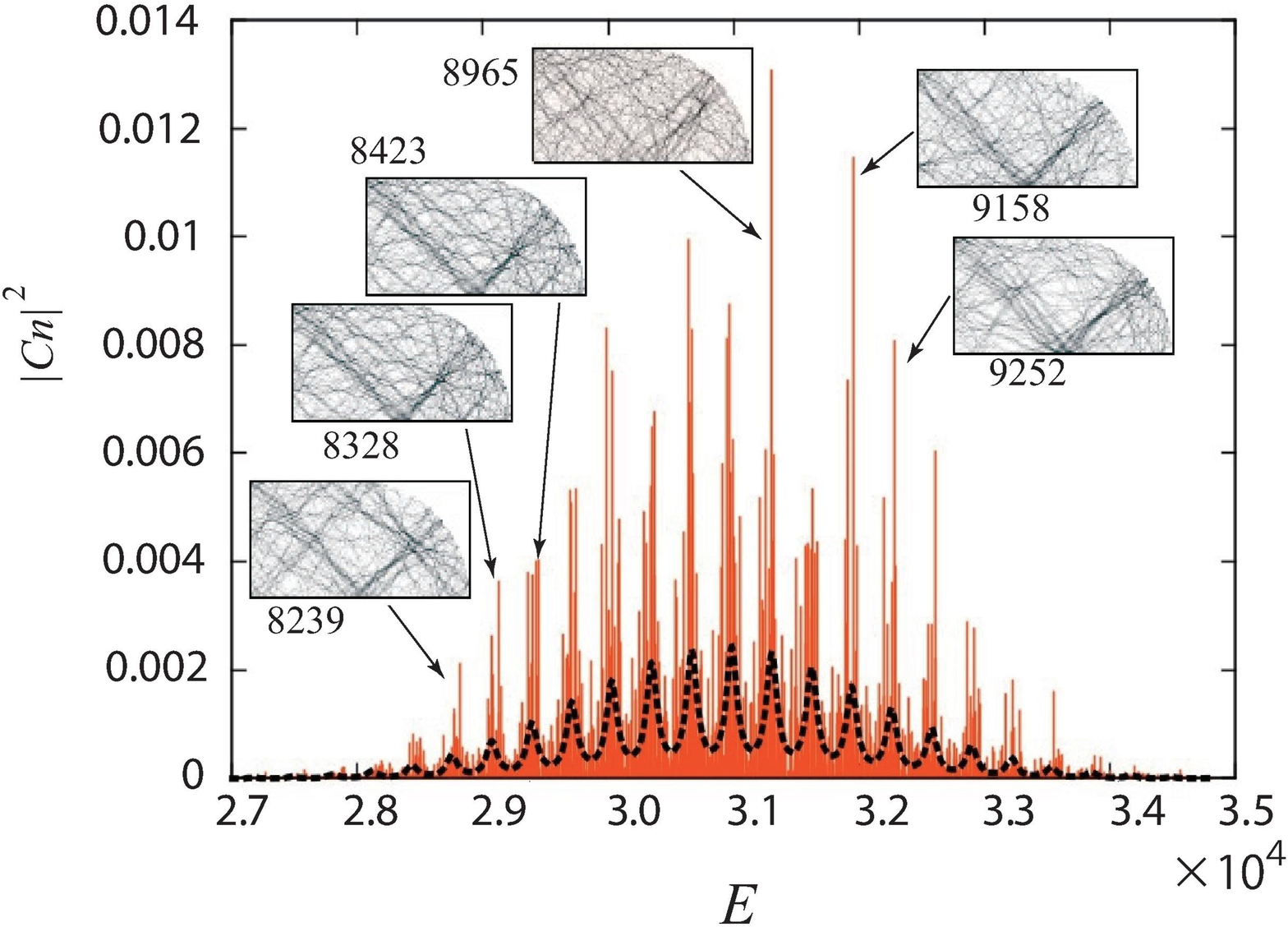}
\caption{Window function (weighted spectrum) of the Gaussian wavepacket $w (E)$ (a dotted curve) for orbit No.7 in Fig.2 is compared with its expansion coefficients $|c_n|^2$ (bars).  Here, the parameters of the initial Gaussian (Eq.(1)) are the same as those in Fig.1.  Insets show the eigen states corresponding to the high peaks.  The 4-digit numbers near the insets represent the counts from the ground state to the excited states in the insets.
``Dynamical scars" are often observed on the classical orbit No.7, as in Fig.2.  
The plot of $|c_n|^2$ is extremly spiky; however, 
it is a typical structure of  the ``totalitarian" case in \cite{KH-LinearNonlinear}. (color online)
\label{fig.4}}
\end{figure}

\begin{figure}
\includegraphics[width=13cm]{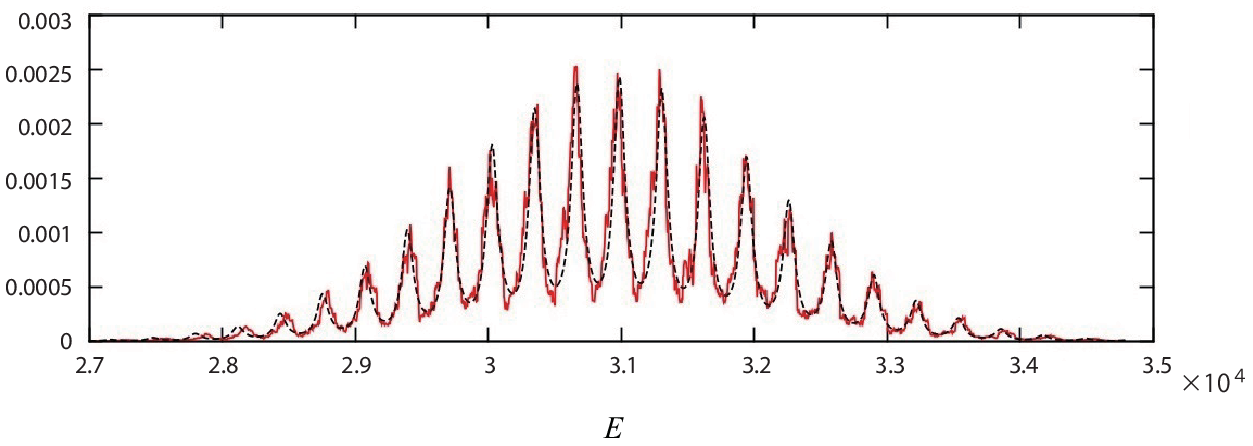}
\caption{Window function (weighted spectrum) of the Gaussian wavepacket $w (E)$  (a dotted curve) for orbit No.7 in Fig.2 and its averaged behavior of the expansion coefficients $|c_n|^2$ (a solid curve).  Here, the averaging is performed in the energy range of $20\epsilon$.  These two lines match very closely. (color online)    
\label{fig.5}}
\end{figure}

In a two-dimensional flat and infinite space, the travelling wavepacket can be calculated exactly. The autocorrelation function is then well approximated as,
\begin{equation}
C_f(t) = \int \Psi_0^* (\mathbf{r} ) \Psi(\mathbf{r},t) d^2 \mathbf{r}  \approx  exp ( - \frac{v^2 t^2}{4 \sigma_0^2}-\frac{i}{\hbar} E_0 t )  ,
\end{equation}
and its real phase part 
\begin{equation}
C_{R}(t) \approx  exp ( - \frac{v^2 t^2}{4 \sigma_0^2} ) 
\end{equation}
satisfactorily represents the damping behavior of the correlation function $C_f (t)$.

In a chaotic finite region, the autocorrelation function should differ as 
\begin{equation}
C(t) \approx \sum_n exp {\left\{ - \frac{v^2 (t-n\tau)^2}{4 \sigma_0^2}-\frac{i}{\hbar} E_0 (t-n\tau) \right\}}
 exp( - \frac{\lambda}{2} |t|) ,
\end{equation}
where $\tau$ is the period of a paticular periodic orbit,
along which the initial wavepacket is launched \cite{LesHouches}.
The summation implies that the finite region allows the wavepacket to repeatedly return to its original location. 
Moreover, its chaoticity makes it spread all over the billiard exponetially under the Lyapnov exponent $\lambda$ of the periodic orbit.
It can be reformed using the Poisson sum rule as
\begin{equation}
C(t)=\sum_{n} \frac{1}{\hbar}  \frac{\Delta}{\sqrt{\pi}}  \frac{\sigma_0}{v}
exp{\left\{ -\frac{\sigma_0^2}{v^2 \hbar^2} (E_n-E_0)^2 \right\}} 
e^{- \frac{i}{\hbar}E_n t} e^{- \frac{\lambda}{2}|t|}   ,
\end{equation} 
where $\Delta=2 \pi \hbar / \tau (=\hbar \omega)$,
$E_n=\Delta n$, and $E_{0}=\frac{\mathbf{{p}_0}^2}{2m}$. 
The weighted power spectrum can then be derived through the Fourier transform of  the autocorrelation function (22) as follows:
\begin{align}
\tilde{C}(E)&=\frac{1}{2 \pi} \int_{-\infty}^{\infty} C(t) e^{\frac{i}{\hbar}Et} dt           \nonumber \\
&=\sum_{n=-\infty}^{\infty} \frac{1}{\hbar} \frac{\Delta}{\sqrt{\pi}} \frac{\sigma_0}{v}
exp{\left\{ -\frac{\sigma_0^2}{v^2 \hbar^2} (E_n-E_0)^2 \right\}}                              \times \frac{1}{\pi} \frac{\lambda/2}{((E-E_n)/\hbar)^2+(\lambda/2)^2}  .
\end{align}
This also includes the Lorentzian function of (12); however, the origin of its peaky behavior is completely different from $\epsilon$.  The Lyapnov exponent $\lambda$ is purely due to the chaotic property of our system and does not exist in $C_f (t)$. 

Therefore, replacing $\tilde{C}_0 (E)$ with $\tilde{C}(E)$, the relation between the window function and power spectrum should be modified to 
\begin{equation}
w(E) \cong - \frac{2\epsilon}{\hbar} \tilde{C}(E)  .
\end{equation}
Then, by Eq.(23), the window function is expected to be
\begin{align}
w (E)  \approx  -2\epsilon & \frac{1}{\sqrt{\pi}} \frac{\sigma_0}{ v }  
exp{\left\{ -\frac{\sigma_0^2}{v^2 \hbar^2} (E-E_0)^2 \right\}}  
\times
  \nonumber  \\
&\times 
\frac{ \Delta}{\pi} \sum_{n=-\infty}^{+\infty} \frac{ \lambda/2}{(E-E_p-n\Delta)^2+(\hbar \lambda/2)^2}.  
\end{align}
Here, $E_p$ represents the energy at the highest maximum of the serial local peaks with width $\lambda$, which is the Lyapnov exponent of the billiard, and $\Delta$ is the energy gap between local peaks.  
The interplay of the Gaussian envelope shape with its width $v \hbar / \sigma_0$ is due to the size of the initial Gaussian (1) and the narrow peaks, with width $\lambda$, represented by the Lorentzian. 
Finally, $w(E)$ is well estimated through Eq.(25) by replacing the eigen energies $E_n$ of the eigenstates in the exponential function of Eq.(23) with an ordinary energy variable $E$.  In reality, the resulting numerical difference of $w(E)$ is slight under the replacement. Then, by using the summation symbol, Eq.(25) simply adds the Lorentzian ``delta" functions, which are smoothed by the Lyapnov exponent $\lambda$.

\begin{figure}
\includegraphics[width=13cm]{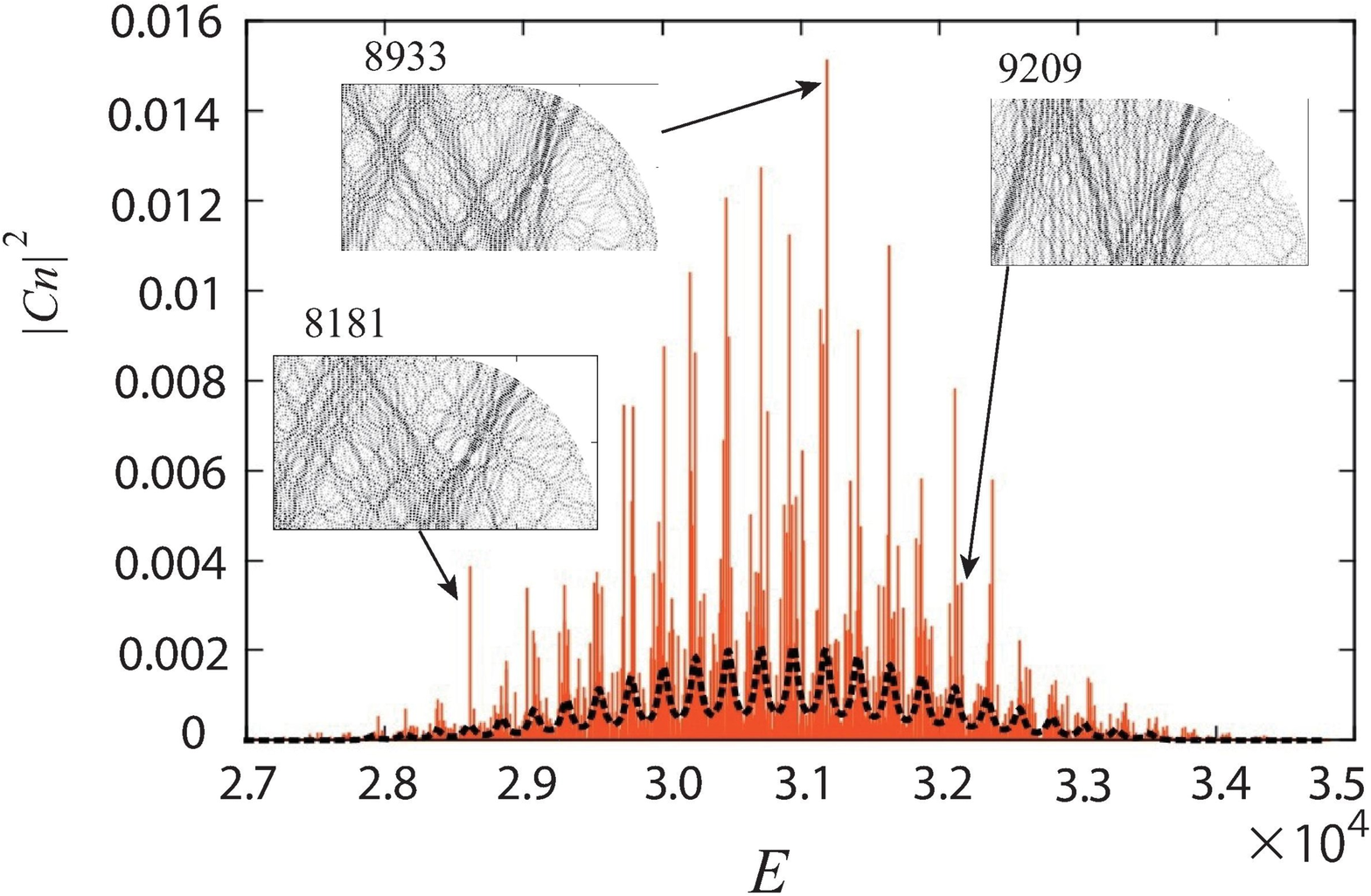}
\caption{
Window function (the weighted spectrum)  of the Gaussian wavepacket $w(E)$ (a dotted curve) for orbit No.14 in Fig.3(b) is compared with its expansion coefficients $|c_n|^2$ (bars).  Here, the parameters of the initial Gaussian (Eq.(1)) are the same as those in Fig.3(b).  Insets show the eigen states corresponding to the high peaks. The 4-digit numbers near the insets represent the counts from the ground state to the excited states in the insets.   ``Dynamical scars" are often present on classical orbit No.14.
The extremely spiky characterictic feature of this $|c_n|^2$ plot is the same as that of No.7 (Fig.4). (color online)    
\label{fig.6}
}
\end{figure}

\begin{figure}
\includegraphics[width=13cm]{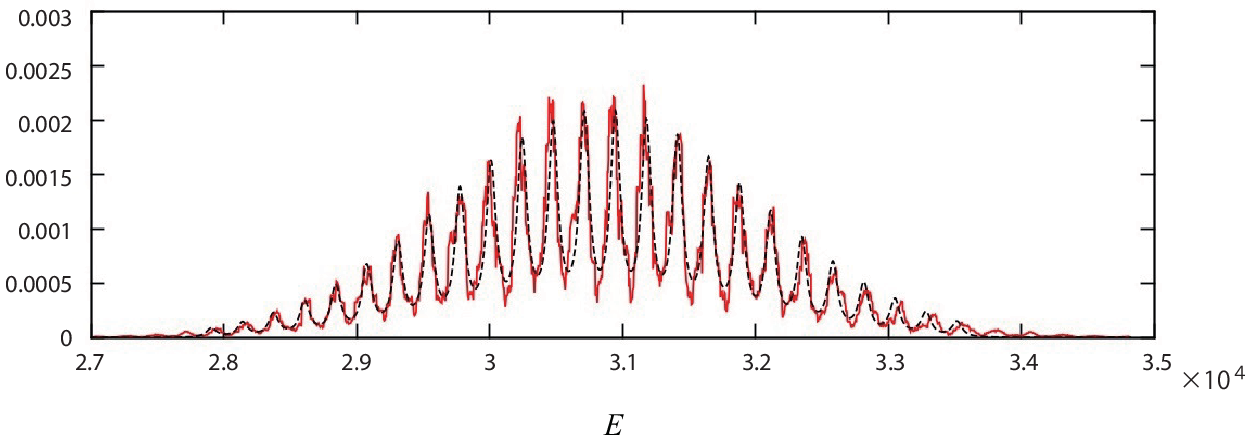}
\caption{Window function (the weighted spectrum)  of the Gaussian wavepacket $w (E)$ (a dotted curve) for orbit No.14 in Fig.3(b), and its averaged behavior of the expansion coefficients $|c_n|^2$(a solid curve).  Here, the averaging is performed in the energy range of $20\epsilon$.  These two lines match very closely. (color online)    
\label{fig.7}}
\end{figure}

In chaotic billiard systems, the actual weighted power spectrum $\tilde{C}(E)$, which is evaluated from numerically obtained eigen states, is known to have an extremely spiky and oscillatory behavior \cite{LesHouches, gaussian, KH-LinearNonlinear, KH-shorttime}. 
The existence of the scar states in chaotic billiard systems leads to a relatively smaller amount of selected eigen states contributing dominantly to $A(\mathbf{r})$.  The $|c_n|^2$ histograms clearly show this tendency. Fig.4 and 6 show the histograms for No.7 and 14 respectively, where the numbering stands for a specific periodic orbit in the stadium, as shown in Table 1 of  \cite{Bogomolny}.

In Fig.4, the red curve represents $w (E)$ for No.7, with $\lambda=0.418|\mathbf{p}_0|$. The constant 0.418 is the geometric Lyapunov exponent and was evaluated from the monodromy matrix of the corresponding periodic orbit \cite{Bogomolny}. In addition, $\epsilon$ is set to the averaged energy level spacing $\Delta E =0.0003412 \times 10^4$.
Other parameters related to the initial Gaussian are the same as those in FIG.1.
They are simply the linear-dynamical predictions of the window function \cite{LesHouches, gaussian, KH-LinearNonlinear, KH-shorttime}.  
The local peaks of the actual weighted spectrum are located at almost equal energy intervals, that is, $\Delta=0.03193 \times 10^4$; this is very close to the theoretical estimation $\Delta_{th}=\frac{\hbar}{m}(\frac{2\pi}{L})|{\mathbf{p}_0|}=0.03253 \times 10^4$, where $L=4.8284$ is the length of the specific periodic orbit.
Through semiclassical approximation, the classical action on the classical periodic orbit is determined as $S_{r}(\xi, \xi; E_0)=\oint_r \mathbf{p} d\mathbf{r} = L \sqrt{2mE_0}$.    
It must increase by as much as $2\pi \hbar$, adding $\Delta_{th}$ to its energy $E_0$.

As aformentioned, $w(E)$ is less spikier than the actual $|c_n|^2$ histogram.  In addition, it is the ``totalitarian" case in Ref. \cite{KH-LinearNonlinear}.  In the weighted spectrum of the "totalitarian" system, some paticular states have dominant contributions. The scars can often be found in such states.  Still, its smoothed behavior follows the estimated envelop function: the window $w(E)$. (The opposite case is called the "egalitarian" in \cite{KH-LinearNonlinear}.  Then the weighted spectrum essentially follows the window function. )  It simultaneously allows the emergence of ``dynamical scars".  Similar to the scar states, if only one primitive periodic orbit has a dominant contribution, the ``dynamical scars" become visible.  
In actuality, the eigen states at peaks often become the scar states of the corresponding periodic orbit (cf. Fig4, 6).  Of course, the eigen states with larger $c_n$ would also contribute to the ``dynamical scars".  However, in some cases, the ``dynamical scars" are blurred by the superposition of the other orbits on the eigen state.

The histogram of $|c_n|^2$s is extremely spiky, although it is possible to enlucidate its smoothed version (Fig.5) formed 
by averaging the energy range, which is sufficiently larger than the energy spacing of levels but much smaller than the required energy. 
It agrees strikingly with the window function $w (E)$.

The same situation occurs for periodic orbit No.14 (Fig3.(b)) in FIG.6, and for orbit No.5, which is already published in \cite{prep}. 
In FIG.6, the red curve represents $w (E)$, with $\lambda=0.3684|\mathbf{p}_0|$. 
The local peaks' energy intervals $\Delta=0.02340 \times 10^4$ are extremely close to its prediction $\Delta_{th}=\frac{\hbar}{m}(\frac{2\pi}{L}){|\mathbf{p}_0|}=0.02428 \times 10^4$ ($L=6.47$).
Moreover, other parameters related to the initial Gaussian are the same as those in Fig.3(b).
In addition, the processes in the smoothed histogram are the same.  The smoothed histogram matches very closely with its window function $w (E)$ (Fig.7).

\begin{figure}
\includegraphics[width=13cm]{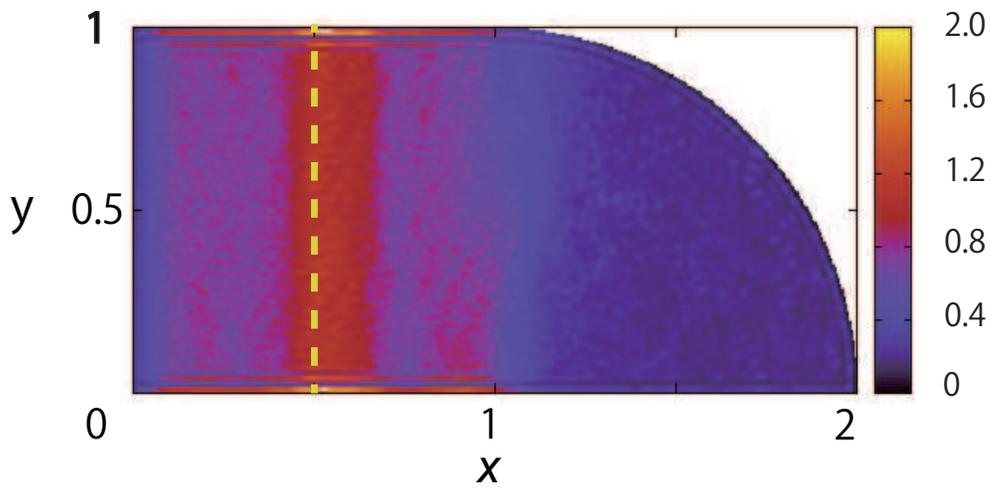}
\caption{The time-average of the evolving wavepacket $A(\mathbf{r})$ on the bouncing ball mode of stadium billiard.  The initial Gaussian wavepacket is set  $|\mathbf{p}_0| = 250$ and $\sigma_0=0.15$.   The wavepacket is launched from$ (1/2, \sqrt{3}/4)$ and the lauching angle is $\theta = \pi / 2$. The broken yellow line corresponds to the classical periodic orbit.  It belongs to the one-parameter family of the bouncing ball mode, whose members bounce up and down between two parallel straight sections of the boundary infinitely, and the launching point is on the line. 
\label{fig.8}}
\end{figure}

\begin{figure}
\includegraphics[width=13cm]{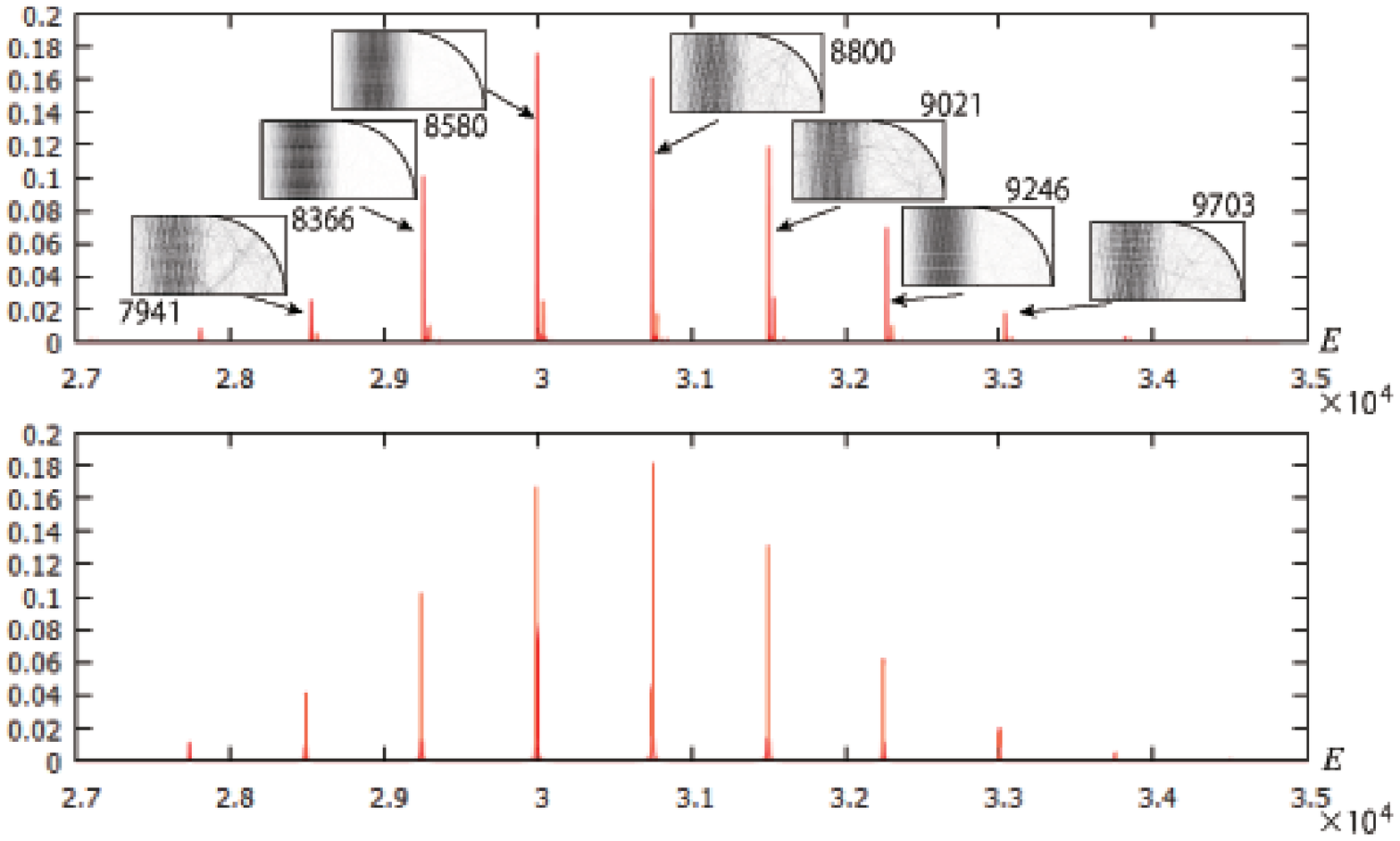}
\caption{Expansion coefficients $|c_n|^2$ (upper graph) and window function (the weighted spectrum) of the Gaussian wavepacket $w (E)$ (lower graph) for the bouncing ball mode. These graphs are almost identical.   The 4-digit numbers near the insets represent the counts from the ground state to the excited states. The parameters of the wavepacket are the same as those in Fig.8. (color online)    
\label{fig.9}}
\end{figure}

Moreover, the bouncing ball mode produces a considerably unique result(Fig.8). This exceptional mode is the only nonchaotic periodic orbit in the stadium billiard.  It has a zero Lyapunov exponent and no chaotic origin because it bounces between the parallel walls of the billiard in terms of classical mechanics.  However, the parameter $\lambda$ still cannot be set to zero or be infinitesimally small in our numerical calculation because the Lorentzian approaches the Dirac delta function in such a limit; this cannot be presented exactly in numerical calculation. Numerical results clarify that only the wave functions with scars on the boucing ball mode significantly contribute to the ``dynamical scar''. Fig.9 compares the numerical histogram and the estimated weighted spectrum, both of which show strikingly good agreement.  Numerically calculated interval between the peaks is $\Delta=0.07524$, whereas its theoretical estimation is  $\Delta_{th}=\frac{\hbar}{m}(\frac{2\pi}{L})|p_0|=0.07854$ ($L=2$).  Note that the width of the sharp peaks $\lambda$ in the weighted spectrum is replaced by averaged level spacing $\overline{\Delta E}$, instead of the theoretically exact value of vanishing Lyapnov exponent $\lambda=0$. 
It also implies that this system does not have much finer energy resolution than $\overline{\Delta E}$.

As mentioned earlier, with a good agreeement between $w(E)$ and the averaged behavior of $|c_n|^2$, the semiclassical approximation can be expected to function satisfactorily in this field.  Moreover, it reminds us of the ``totalitarian" aspect of the system. 

If we choose a sufficiently small window size to reasonably suppose that only one eigen state would be in the window simultaneously, it essentially resembles the result of Ref.\cite{Bogomolny} for the scar states.  However, in this study the window size is much larger because the initial wavepacket must involve the contribution of eigen states in a broader energy range.  
Thus, a scar is not directly observed in the snapshot of time-dependent wave functions (Fig.1(f)).  The ``dynamical scar" is the superposition of many corresponding states in the energy window.

\begin{figure}
\includegraphics[width=13cm]{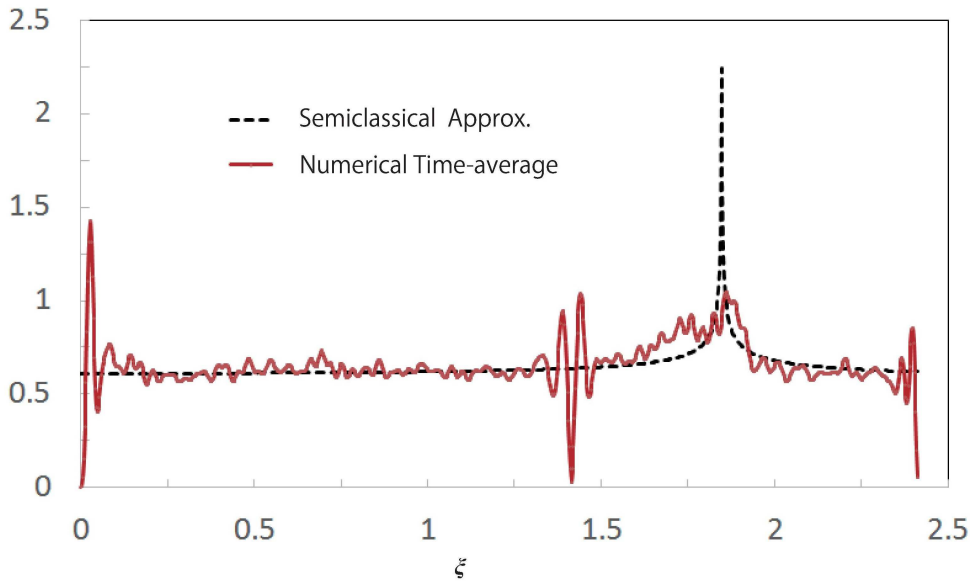}
\caption{Comparison of the semiclassically approximated time-average of evolving wavepacket (29) on periodic orbit No.7(a dotted curve) and its numerically calculated localization(a solid curve).  They are presented as functions of the distance $\xi$ from the point (0,1), which is measured along the broken yellow line in Fig.2.  At the distance $\xi_C=\sqrt{2+\sqrt{2}}=1.8478...$, the semiclassical approximation diverges.  At distances $0$, $\sqrt{2}=1.4142...$ and $1+\sqrt{2}=2.4142..$, the boundary walls are present.   At the boundary, the wave function becomes zero and shows a peculiar rough wavy behavior. (color online)
}
\end{figure}

\section{Semiclassical approximation}

Through  semiclassical approximation \cite{Bogomolny}, 
the localization becomes the summation of  two parts: 
\begin{equation}
A(\mathbf{r})   \cong \langle \rho_0 (\mathbf{r}, E) \rangle 
+ \int w(E) 
\rm{Im} 
\it{G}_{osc}(\mathbf{r},\mathbf{r};E)dE
=\langle \rho_{\rm{0}} (\mathbf{r}, E) \rangle+A_{osc}(\mathbf{r}) , 
\end{equation}
where
\begin{align}
G_{osc}&(\mathbf{r},\mathbf{r};E) \cong \frac{2}{(2 \pi)^{1/2} \hbar^{3/2}} 
\times
\nonumber \\
&\times 
\sum_{\gamma, n} \frac{D_{\gamma,n} (\xi)^{1/2}}{v} 
\left\{
 exp \left[ \frac{i}{\hbar} (S_{\gamma, n}(\xi, \xi; E) + \frac{W_{\gamma,n} (\xi) }{2}\eta^2) \right] 
 - i \frac{\pi \nu_{\gamma,n} }{2} -i \frac{3}{4}\pi
\right\} .
\nonumber \\
\end{align}
The first term of Eq.(26) in the right-hand side is the smooth part $ \langle \rho_0 \rangle $, and the second is the oscillatory term $A_{osc}$.
Further the angle branckets $\langle \cdots \rangle $ denote an average over the energy range that the window function $w(E)$ covers, and $ \rho_0 (\mathbf{r}, E) $ is the classical probability density of finding a particle with energy $E$ at point $\mathbf{r}$.  Needless to say, $w(E)$ depends on the shape of the (initial) wavepacket.
The $\xi$ axis is set along the concerned periodic orbit, and the $\eta$ axis perpendicular to it at point $\xi$.  The classical action  
of the $n$-fold repeated orbit can be derived as $S_{\gamma,n}=nS_\gamma$ from the action of the primitive orbit $\gamma$: $S_\gamma$. 
Then, $T_{\gamma,n} (\mathbf{r},E)=nT_\gamma$, $T_\gamma$ is the period of the primitive orbit $\gamma$.  Its maximal number of conjugate points $\nu_{\gamma,n}=n\nu_\gamma$  can be derived from the primitive $\nu_\gamma$.  
In addition, 
$W_{\gamma,n} (\xi)$, $D_{\gamma,n} (\xi)$ 
are  versions for the $n$-fold periodic orbit and can be expressed by
 $D_\gamma=-(\frac{\partial^2 S_\gamma}{\partial \eta' \partial \eta''})_{\eta'=\eta''=0}$ 
and $W_\gamma(\xi)=(\frac{\partial^2 S_\gamma}{\partial \eta'^2} + \frac{\partial^2 S_\gamma}{\partial \eta' \partial \eta''} + \frac{\partial^2 S_\gamma}{\partial \eta''^2})_{\eta'=\eta''=0}$ for the primitive orbit. They 
can be derived from $D_\gamma$:
$D_{\gamma,n}(\xi)=D_\gamma \frac{\mu_1 - \mu_2}{\mu_1^n - \mu_2^n}$, $W_{\gamma,n}(\xi)=D_{\gamma,n}(\mu_1^n + \mu_2^n - 2)$.  
Note that $\mu_1$, $\mu_2=\mu_1^{-1}$ are the eigenvalues of the monodromy matrix of the primitive orbit.  

It is assumed that only one specific periodic orbit $\gamma=C$ shows a prime contribution.  Moreover, primitive orbit $n=1$ is expected to be dominant on the periodic orbit because the factor $D_{C,n}$ vanishes rapidly with increasing $n$.  Therefore, the oscillatory part of $A$ can be approximated on the classical orbit $C$ ($\eta=0$) as 
\begin{align}
A_{osc}&(\xi) \cong  
  \frac{2\sqrt{2}}{\pi \hbar^{7/2}} \frac{\sigma_0}{v} \epsilon \Delta
\sum_{j} exp[- \frac{{\sigma_0}^2}{\hbar^2 v^2}(E_j - E_0)^2 ] \frac{|D_{C}|^{1/2}}{v} \times
\nonumber \\
&\times
\int \frac{1}{\pi} 
\frac{ \lambda /2}{ \{(E-E_j)/\hbar\}^2 + (\lambda/2)^2} 
Im 
\{ i \exp [\frac{i}{\hbar} S_{C} -i \frac{\pi}{2}\nu_C + i \pi N_C - i \frac{1}{4} \pi ] \}
 dE .
 \nonumber \\
\end{align} 
Note that $N_C$ is the number of hits on the boundary, when a particle travels around the closed orbit $C$, and $D_C = D_{C,1}$.  Under the semiclassical approximation, at $E=E_j$, 
it can be well assumed that
 $exp \{ \frac{i}{\hbar} S_{C}(\xi,\xi;E_j)-i \frac{\pi}{2} \nu_C + i \pi N_C -i \frac{1}{4} \pi \} =1 $.
Finally, the integration in Eq.(28) can be performed using the complex integral, and the localization is evaluated as
\begin{align}
A(\xi)&=\langle \rho \rangle + A_{osc}(\xi, E)
\nonumber \\
&= \frac{1}{Area} + \frac{2\sqrt{2}}{\pi \hbar^{5/2}} \frac{\sigma_0}{v} \epsilon \Delta \frac{|D_{C}(\xi)|^{1/2}}{v}
\sum_j exp [ -\frac{ {\sigma_0}^2 }{\hbar^2 v^2} (E_j - E_0)^2 ] e^{-T_j \frac{\lambda}{2} }
 \nonumber \\
\end{align}
where $S_{C}(\xi, \xi;E_j + i \frac{\hbar \lambda}{2} ) \cong S_{C}(\xi, \xi;E_j)+i T_j \frac{\lambda \hbar}{2} $ is used, $T_j$ is the period of the periodic orbit at $E=E_j$, and $Area$ is just the area of the billiard.  
Finally, the averaged level spacing $\overline{\Delta E}$, which is the criterion of the energy resolution limit of the billiard system, is adopted for $\epsilon$

\begin{figure}
\includegraphics[width=13cm]{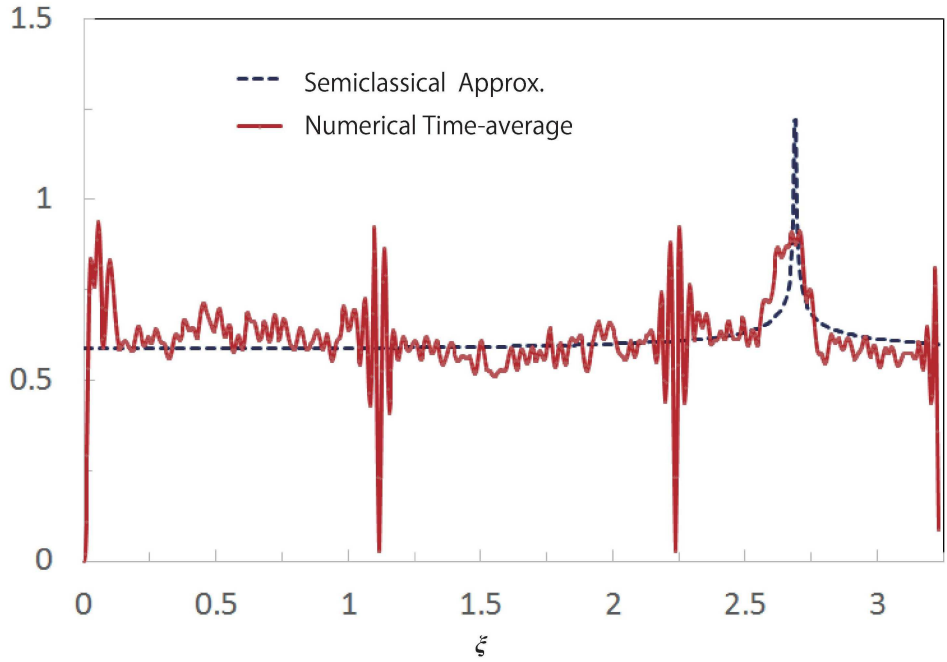}
 \caption{Comparison of the semiclassically approximated time-average of evolving wavepacket (29) on periodic orbit No.14(a dotted curve) and its numerically calculated localization(a solid curve).  They are presented as functions of the distance $\xi$ from the point (0,0), which is measured along the broken yellow line in Fig.3(b).  At the distance $\xi_C=\sqrt{5+\sqrt{5}}=2.6900...$, the semiclassical approximation diverges.  At distances $0$, $\frac{\sqrt{5}}{2}=1.1180...$, $\sqrt{5}=2.2361...$, and $1+\sqrt{5}=3.2361...$, the boundary walls exist.   At the boundary, the wave function becomes zero and shows a peculiar rough wavy behavior. (color online)
\label{fig.11}}
\end{figure}

The evaluated localization $A$ on the periodic orbit No.7(FIG.2) is presented in Fig.10.  Assuming the wave function is completely flat in the finite region,  $\langle \rho  \rangle $ must be the inverse of the area of the billiard: $ \lbrace (4+\pi)/4 \rbrace ^{-1} =0.5601...$ throughout the stadium. 
Owing to the scar or the contribution of the classical periodic orbit, the concentration enhances the absolute square of the wave function by at least $10\%$ on the periodic orbit above the average behavior $\langle \rho  \rangle $, except in the neighborhood of the singularity around the conjugate point.  
Of course, it cannot recreate the wavy behavior, which is especially sharp close to the boundary because Eq.(29) does not show the exact effect of the boundary condition.  The approximation is determined essentially through the length of the orbits and the energy.  Actually, the wave must be zero at the boundary according to the Dirichlet condition, and all dominant eigenfunctions' phases become almost coherent near the boundary.  
Fig.11 shows the semiclassical approximation of No.14. In addition, it presents essentially the same results as No.7 (Fig.10).

The singularity at the conjugate point is inevitable for the semiclasical approximation; however, it is also beyond the scope of the approximation in the neighborhood of the point. The semiclassical approximation of the wave function diverges at the point due to factor $D_C = 1/m_{12}$, and $m_{12}$ is the off-diagonal element of the monodromy matrix\cite{Bogomolny} for the unstable classical periodic orbit $C$.  In our study, $m_{12}=-2\{(2+\sqrt{2})-\xi^2\}$ for No.7 (Fig.10), and  $m_{12}=-2\{(5+\sqrt{5})-\xi^2\}$ for No.14 (Fig.11).  In both cases $\xi$ is measured from the left wall and along the orbits.  The monodromy matrix element $m_{12}$ becomes zero and $D_C$ diverges at the conjugate point $\xi_{C}$, where the classical orbits near the classical periodic orbit converge. The conjugate points are located at $\xi_{C}=\sqrt{2+\sqrt{2}}$ for No.7 mearsured from the point $(0,1)$, and $\sqrt{5+\sqrt{5}}$ for No.14 measured from $(0,0)$.   In reality, a relatively strong enhancement exists around the point.
Apart from these properties,  the semiclassical approximation works well, and Eq.(29) still matches remarkably with the numerically evaluated time-averages on the orbits.

\section{Conclusion}

The quantum phenomenon: the ``dynamical scar" is analyzed from the aspect of the eigen state expansion of the incident wavepacket and the semiclassical approximation.  By launching a Gaussian wavepacket along a classical unstable periodic orbit, its weighted power spectrum $\bar{C}(E)$ accomplishes a good match with its averaged histogram of expansion coefficients $|c_n|^2$s.  

By utilizing $\bar{C}(E)$ as the energy window function for the semiclassical approximation, the ``dynamical scars" can be evaluated.  The periodic orbit critically contributes to the approximation.  However, it has nonrealistic singularities close to the conjugate points on the orbit.
The window function $w(E)$, which is manipulated from $\tilde{C}(E)$, plays a crucial role for the approximation.

By setting the window size small so that only one eigen state can exist inside the window energy range, our discussion then becomes the same as the scar state theory of Bogomolny \cite{Bogomolny}.  In this study, the window size was sufficiently large to include more than several scarred eigen states to make the ``dynamical scar" clearly visible.  Simultaneously, this may be why we cannot observe scars in the snapshots of traveling wave functions after their diffusing throughout the billiard (Fig.1(f)).   The ``dynamical scar" is the interplay of many related scarred states inside the range of the energy window.


\begin{thebibliography}{00}


%%\bibitem{}

\bibitem{heller}
{	E. J. Heller,
     Bound-State Eigenfunctions of classically Chaotic Hamiltonian Systems: Scars of Periodic Orbits,
	Phys. Rev. Lett.
	{\bf 53},
	(1984),
	1515-1518.
}

\bibitem{Bogomolny}
{	E. B. Bogomolny,
    Smoothed Wave Functions of Chaotic Quantum Systems,    
	Physica D
	{\bf 31},
	(1988)
	169-189.
}

\bibitem{LesHouches}
{
	E. J. Heller,
	\textit{Chaos and Quantum Physics},
	edited by M. J. Giannoni, A. Voros, and J. Zinn-Justin,
	Les Houches Session LII, 1989
	Elsevier, Amsterdam, 1991
	pp.547-663.
}

\bibitem{Berry-Wignerdist}
{	M. V. Berry,
    Quantum Scars of Classical Closed Orbits in Phase Space,
	Proc. R. Soc. Lond. A
	{\bf 423},
	(1989)
	219-231.
}


\bibitem{heller2}
{	E. J. Heller,
    Quantum Localization and the Rate of Exploration of Phase Space,
	Phys. Rev. A
	{\bf 35},
	(1987)
	1360-1370.
}

\bibitem{TH}
{
	S. Tomsovic and E. J. Heller,
     Semiclassical Construction of Chaotic Eigenstates,
	Phys. Rev. Lett.
	{\bf 70},
	(1993)
	1405-1408.    
}

\bibitem{gaussian}
{
	S. Tomsovic and E. J. Heller,
    Long-time Semiclassical Dynamics of Chaos: the Stadium,
	Phys. Rev. E
	{\bf 47},
	(1993)
	282-299.
}

\bibitem{KH-LinearNonlinear}
{    L. Kaplan and E. J. Heller,
     Linear and Nonlinear Theory of Eigenfunction Scars,
     Ann. Phys. (N.Y.)
     {\bf 264},
     (1998)171-206.
}

\bibitem{KH-shorttime}
{    L. Kaplan and E. J. Heller,
Short-time Effects on Eigenstate Structure in Sinai Billiards and Related Systems,
     Phys. Rev. E
     {\bf 62},
     (2000)
     409-426.
}


\bibitem{ourpaper} 
{
	H. Tsuyuki, M. Tomiya, S. Sakamoto and M. Nishikawa,
	Scar-Like States in Dynamical Electron-Wavepackets in Chaotic Billiard,
    e-J. Surf. Sci. Nanotech.
	{\bf 7},
	(2009)721-727.
%	https://www.jstage.jst.go.jp/article/ejssnt/7/0/7_0_721/_pdf
}


\bibitem{ourpaper2}
{
	M. Tomiya, H. Tsuyuki and S. Sakamoto,
     Quantum Fidelity and Dynamical Scar States on Chaotic Billiard System,
	Comm. Comp. Phys.
	{\bf 182},
	(2011)
	245-248.
}


\bibitem{prep}
{   M. Tomiya, H. Tsuyuki, K. Kawamura, S. Sakamoto and E. Heller,
     Scar State on Time-evolving Wavepacket,
    J. Phys. Conf. Ser. 
    {\bf 640},
    (2015)012068.
}



\bibitem{St}
{    H-J. St\"{o}ckmann,
     \textit{Quantum Chaos an Introduction}
     Cambridge University Press, Cambridge, 1999,
     pp.305-310.
}    


\bibitem{Bunimovich}
{	L. A. Bunimovich,
    On Ergotic Properties of Certain Billiards,
	Funct. Anal. Appl.
	{\bf 8},
	(1974)
	254-255.
}


\bibitem{Berry85}
{
	M. V. Berry,
	Semiclassical Theory of Spectral Rigidity,
	Proc. R. Soc. Lond.
	A {\bf 400},
	(1985)
	229-251.
}


\end{thebibliography}
\end{document}